\documentclass[11pt,a4paper]{article}
\usepackage[T1]{fontenc}
\usepackage{iftex}
\ifPDFTeX
\usepackage[utf8]{inputenc}
\fi
\usepackage{lmodern}
\usepackage[margin=25mm]{geometry}
\usepackage{amsmath,amssymb,graphicx,booktabs,array,tabularx}
\usepackage{caption}
\usepackage[hidelinks]{hyperref}
\usepackage{microtype}
\title{Residual-Guided Multi-Step Conditional Generation for Fast Airfoil Flow Field Prediction}
\author{Weishao Tang\textsuperscript{1}, Mingcheng Lei\textsuperscript{1}, and
Yufei Zhang\textsuperscript{1,2,3}\thanks{Corresponding author: \href{mailto:zhangyufei@tsinghua.edu.cn}{zhangyufei@tsinghua.edu.cn}}\\[6pt]
\small\textsuperscript{1}School of Aerospace Engineering, Tsinghua University, Beijing, China\\
\small\textsuperscript{2}State Key Laboratory of Advanced Space Propulsion,\\
\small Tsinghua University, Beijing, China\\
\small\textsuperscript{3}Caofeidian Laboratory, Tangshan, China}
\date{}
\begin{document}
\maketitle
\begin{abstract}
Deep-learning surrogate models for aerodynamic flow prediction are efficient within the training distribution, but their accuracy often degrades sharply in out-of-distribution (OOD) scenarios. To improve OOD robustness for airfoil flow-field inference, this study proposes a residual-guided sequential inference framework to correct the unreliable local flow field. The framework consists of three Transformer-based components: a Flow Initializer for direct flow prediction, a Residual-based Corrector for high-residual regions, and a Shock-based Corrector for shock regions. The framework is evaluated on an engineering-oriented OOD test set built from CFD-based aerodynamic optimization trajectory. The proposed framework reduces the L1 error, CL error, and CD error by 30\%, 59\%, and 41\%, respectively, relative to the usual single step Geo-to-Flow prediction. The additional iterative inference cost is only 0.25s per sample. These results demonstrate that the proposed framework provides an effective route for improving OOD robustness in airfoil surrogate modeling.

\end{abstract}
\section{Introduction}

Computational Fluid Dynamics (CFD) based on traditional numerical methods is computationally expensive for tasks such as large-scale three-dimensional aerodynamic optimization \cite{ref1}. Recently, several studies have constructed efficient surrogate models for aircraft using artificial intelligence (AI) to achieve high-performance simulation in aerodynamic design \cite{ref2,ref3}. However, a significant challenge for current AI methods in engineering tasks is that while models infer efficiently and accurately within the training distribution, their prediction errors often increase sharply and uncontrollably in out-of-distribution (OOD) scenarios \cite{ref3}. In practical engineering, OOD scenarios are unavoidable, as it is difficult to sufficiently sample all possible geometries of interest during the dataset construction phase, especially for higher-dimensional 3D problems. As a result, repeated CFD simulations on a large number of OOD geometric samples and continuous fine-tuning of the model are required to ensure model accuracy, which in turn increases the application complexity of the deep learning surrogate model framework. Therefore, enhancing the OOD generalization performance of AI surrogate models is critical to improving their engineering utility.

In recent years, some work in the field of AI has sought to improve OOD generalization by integrating numerical solvers with neural networks, enforcing physical constraints during flow field inference. For instance, Yang et al. incorporated the residual of the governing equations into the training loss to guide the model in satisfying physical conservation laws during training, which improves the out-of-distribution physical robustness of the model \cite{ref4}. ANCHOR \cite{ref5} invokes traditional numerical methods within the inference framework when residuals exceed a specific threshold, achieving stable predictions for 3D unsteady diffusion equations. PIRF \cite{ref6} serializes complex mappings and utilizes equation residuals to constrain generated results in real-time during inference, thereby improving OOD predictions. However, these approaches have not yet been applied to fluid dynamics problems characterized by stronger nonlinearities.

Inspired by these ideas, this study plans to construct an innovative fast inference framework that fuses traditional numerical methods with deep learning, applied to airfoil flow field inference to enhance OOD generalization. We construct an OOD test set based on uncontrolled airfoil geometries collected during the aerodynamic optimization process to simulate the scenario of applying deep learning models to practical engineering. The following sections will respectively introduce the dataset construction scheme, the constructed multi-stage alternating flow field correction operator, and the framework's performance on the OOD test set.

\section{Dataset Establishment}

This section introduces the construction of the training dataset and the out-of-distribution (OOD) test dataset.

\begin{figure}[htbp]
\centering
\includegraphics[width=\linewidth]{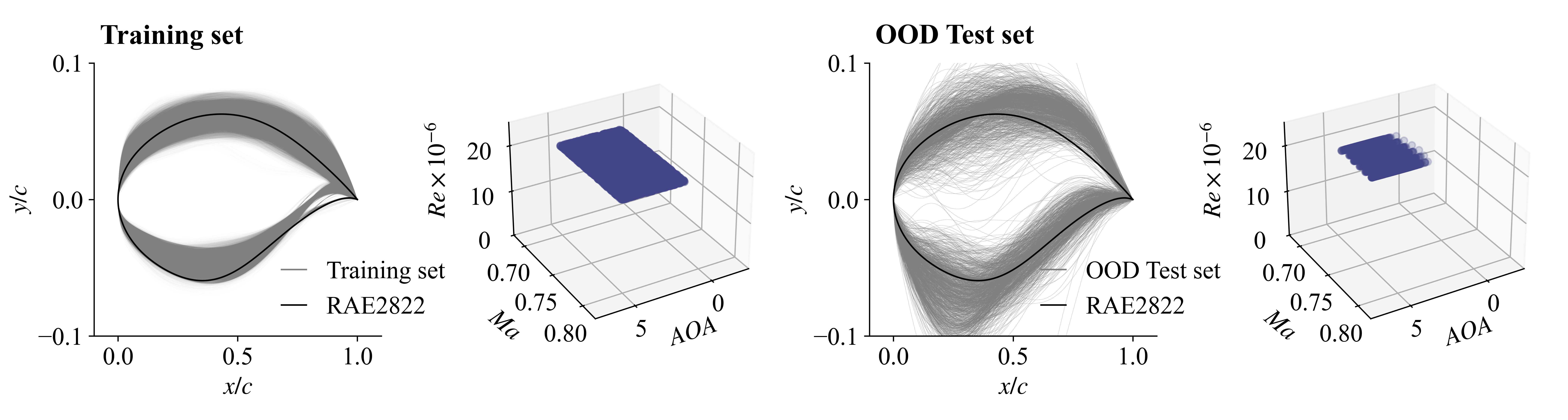}
\caption{The geometry and inflow distribution of training and OOD test set}
\label{fig:1}
\end{figure}

\begin{table}[htbp]
\centering
\caption{The inflow distribution of training and OOD test set}
\label{tab:1}
\small
\begin{tabular}{lll}
\toprule
 & Training Dataset & OOD Test Dataset \\
\midrule
AOA & $\sim U[0,5]$ & $\sim C_L(0.824)$ \\
Ma & $\sim U[0.70,0.80]$ & $0.71,0.72,0.73,0.74,0.75,0.76$ \\
$\mathrm{Re}\times 10^{-6}$ & 20 & 20 \\
\bottomrule
\end{tabular}
\end{table}

\subsection{Training Dataset Establishment}

The training dataset is constructed using the OSS sampling strategy, which enables the dataset to cover diverse flow structures \cite{ref7}. The final dataset contains 45,000 airfoil samples and their corresponding high-fidelity flow fields, with the geometric distribution covering a representative design space of supercritical airfoils.

The detailed distributions of geometry and freestream conditions are shown in Fig.~\ref{fig:1} and Table~\ref{tab:1}. The dataset is built on structured C-type grids. The grids are generated in batch mode using an in-house script, and the high-fidelity flow fields are computed using the open-source solver CFL3D \cite{ref8}. To reduce the grid-processing overhead in subsequent surrogate-model inference, the flow fields on the original CFD grids are further interpolated onto uniformly defined algebraic grids, and the surrogate models are directly executed on these algebraic grids.

\subsection{Out-of-distribution Test Dataset}

Most existing data-driven flow-field surrogate models are evaluated on in-distribution test sets. To thoroughly assess model performance under realistic engineering OOD scenarios, this study constructs an OOD test dataset based on an aerodynamic optimization task. Specifically, starting from the baseline airfoil RAE2822, the following aerodynamic optimization problem is solved using CFD.

\[
\begin{aligned}
\min_{\mathrm{CST}_{ui},\mathrm{CST}_{li}}\quad
 C_{D,\mathrm{Avg}} &= \frac{1}{6}\sum_{k=1}^{6} C_{D,\mathrm{point}_k} \\
1 \leq \mathrm{AOA}_{\mathrm{point}_k} &\leq 5,\quad k=1,2,\ldots,6 \\
\mathrm{Ma}_{\mathrm{point}_k} &= 0.71,0.72,\ldots,0.76 \\
C_{m,\mathrm{point}_k} &\geq -0.092,\quad k=1,2,\ldots,6 \\
\mathrm{Aera}_{\mathrm{foil}} &\geq \mathrm{Aera}_{\mathrm{foil}_0},\quad
t_{0.15} \geq 0.9t_{0.15,\mathrm{foil}_0}
\end{aligned}
\]

During the aerodynamic optimization process, all candidate airfoils along the optimization trajectory, together with their corresponding operating conditions, are collected. Since the aerodynamic optimization does not explicitly constrain the airfoil geometry, the trajectory naturally generates a wide variety of geometries outside the training distribution. The resulting geometry and freestream distributions are shown in Fig.~\ref{fig:1} and Table~\ref{tab:1}. This construction gives the OOD test set a clear engineering meaning and allows a rigorous evaluation of model robustness under realistic out-of-distribution conditions.

\section{Method}

This section presents the proposed residual-guided airfoil surface-flow sequence inference framework in detail. The complete framework consists of three components: a Flow Initializer that provides the initial flow field, a Residual-based Corrector that refines generic high-residual regions, and a Shock-based Corrector that further improves shock and strong-discontinuity regions. All three components share a Conditional Transformer backbone and differ only in the construction of training inputs and outputs and in their masking strategies. This section is organized into three subsections, introducing the backbone architecture, the training strategies of the three models, and the final iterative inference framework, respectively.

\subsection{Backbone Network Architecture}

All three components are built based on a near-wall surface Conditional Transformer \cite{ref9}. The target flow field is defined on a common grid layout and contains $(p,T,u,v)$ on 296 chordwise cells and two wall-normal cell layers. The use of two wall-normal cell layers is intended to provide the necessary boundary information for residual evaluation.

\begin{figure}[htbp]
\centering
\includegraphics[width=\linewidth]{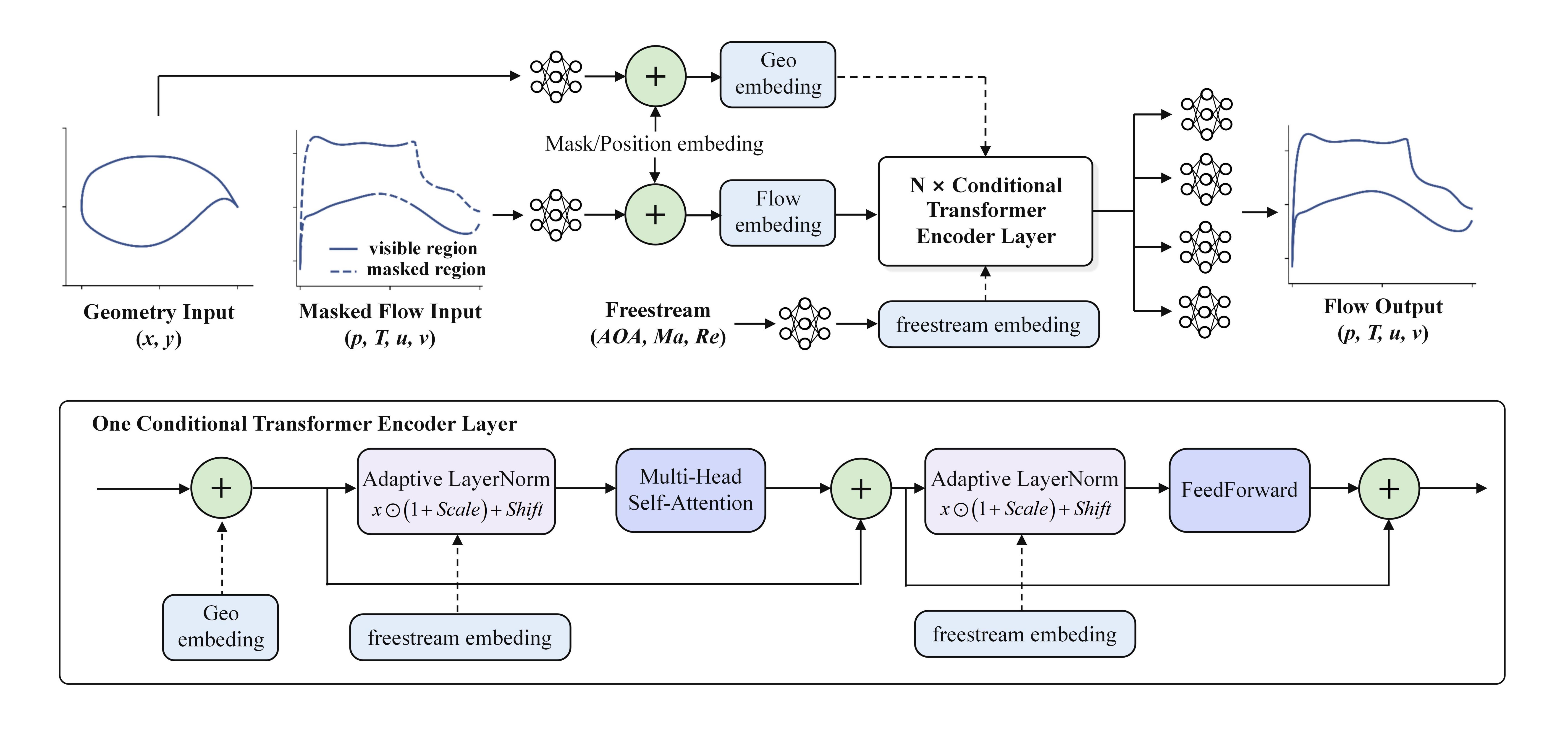}
\caption{Model architecture of the backbone Transformer}
\label{fig:2}
\end{figure}

The overall architecture is shown in Fig.~\ref{fig:2}. The model consists of multiple Transformer blocks with standard self-attention. In each encoder layer, geometry features are re-injected once more so that the model remains focused on the geometric condition and the geometric constraint is not gradually diluted as the depth increases.

The freestream conditions $(\mathrm{AOA},\mathrm{Ma},\mathrm{Re})$ are introduced through an adaptive feature-wise scaling and shifting of a global latent condition. The adopted AdaNorm formulation can be written as

\[
\operatorname{AdaNorm}(x,c)=\operatorname{LayerNorm}(x)
\bigl(1+\operatorname{scale}(c)\bigr)+\operatorname{shift}(c).
\]

where $x$ denotes the hidden representation and $c$ denotes the encoded freestream condition.

\subsection{Model Training Strategies}

\begin{figure}[htbp]
\centering
\includegraphics[width=\linewidth]{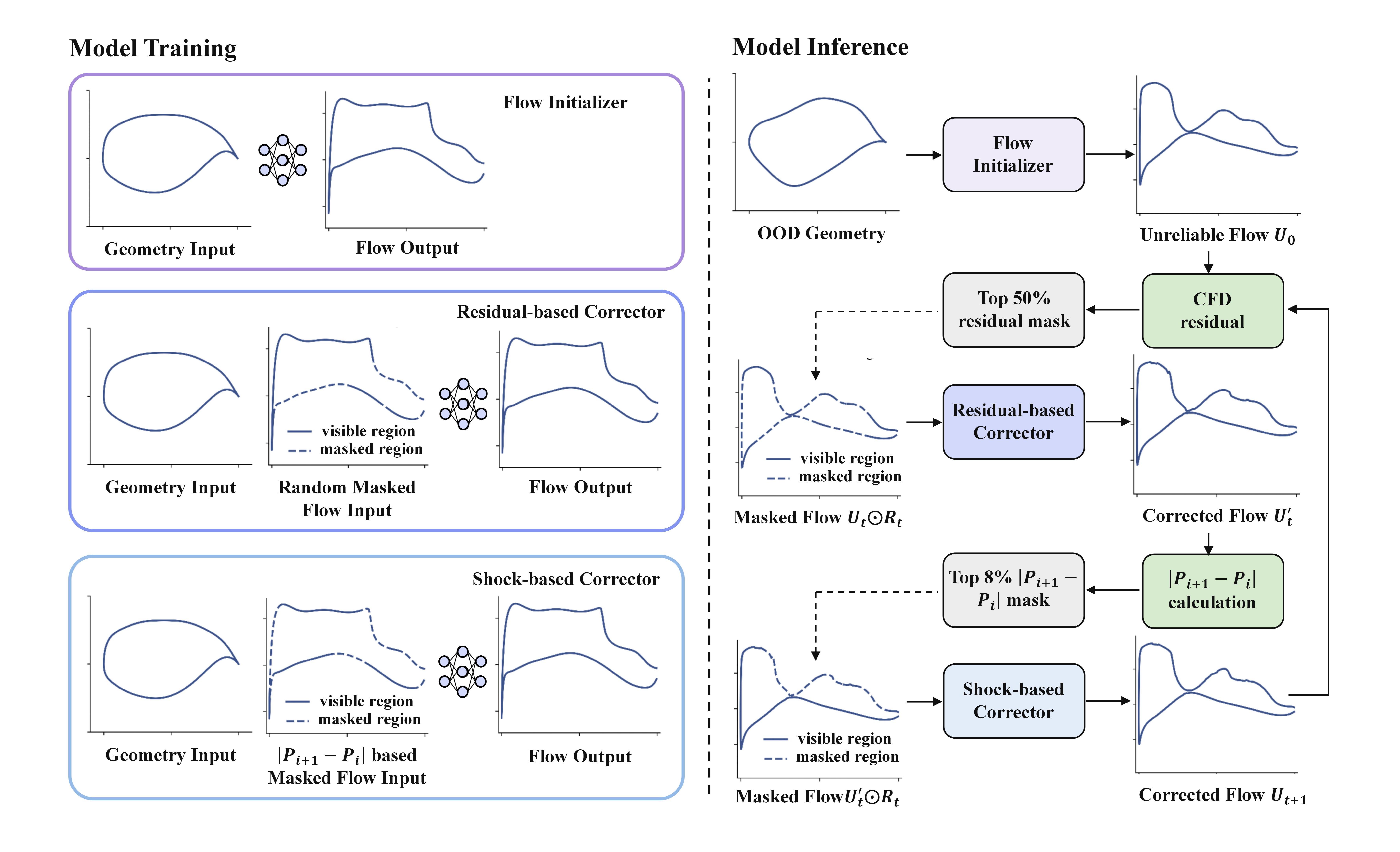}
\caption{The training and inference framework of the proposed method}
\label{fig:3}
\end{figure}

Although the three models share the same Conditional Transformer backbone, their training tasks are constructed differently, as illustrated in Fig.~\ref{fig:3}. The Flow Initializer directly predicts the near-wall flow field from airfoil geometry and freestream conditions like most existing surrogate models. The Residual-based Corrector and the Shock-based Corrector are established inspired by \cite{ref10,ref11}. They take both flow and geometry as inputs, but only part of the flow field is visible, whereas the geometry remains fully visible. The models reconstruct the masked regions of the flow field from these inputs. In the masked regions, the original flow features are replaced with a learnable mask token.

The Residual-based Corrector and the Shock-based Corrector use different masking strategies. The Residual-based Corrector applies piecewise random masks to the input flow field. The masking ratio is uniformly sampled \textasciitilde{} U[0.4,0.6], and the number of masked segments is sampled \textasciitilde{} U[5,20]. A random mask is then constructed according to the sampled configuration. During training, the reconstruction loss is computed only on the masked tokens, while the unmasked region is excluded from supervision. This design explicitly trains the model to recover the masked flow region using the visible random flow segments together with the complete geometric information.

Unlike the random masking used for the Residual-based Corrector, the Shock-based Corrector focuses primarily on reconstructing strong-discontinuity regions. Specifically, the pressure jump $\lvert p_i-p_{i+1}\rvert$ is first computed on the first wall-normal cell layer. The top 8\% region according to this pressure-jump indicator is selected and then expanded by 8 cells along the chordwise direction to form a shock-sensitive mask. In addition, a small proportion of random masking is superimposed so that the model retains the ability to reconstruct arbitrarily masked flow fields.

\subsection{Alternating-Operator Iterative Inference Framework}

During inference, the above models are organized into a two-stage interleaved iterative framework that starts from an initial flow field. A schematic of the inference framework is shown in Fig.~\ref{fig:3}. First, the Flow Initializer provides the initial flow field for the iterative process. For in-distribution test samples, the accuracy of this initial field is often sufficient. However, for samples outside the training distribution, its predictions are often unreliable. Therefore, the Residual-based Corrector and the Shock-based Corrector are invoked alternately to refine the flow field until the total residual increases, at which point the iteration stops.

In each iteration step, the near-wall residual field is evaluated before calling the Residual-based Corrector. The top 50\% of cells in the residual field are selected as the masked region. The unselected low-residual region is regarded as the relatively reliable part of the flow field and is kept visible as the conditional input to the model. Based on this input, the high-residual region is reconstructed. Since the low-frequency learning bias of neural networks often leads to poor approximation in high-frequency shock regions, a second refinement stage is introduced after the Residual-based Corrector. Specifically, the top 8\% pressure-jump region is selected, and each selected span is expanded by 8 cells on both sides. The resulting masked region is then reconstructed by the Shock-based Corrector using the non-shock region as the visible condition.

After each complete iteration, a stopping criterion is evaluated. If the residual increases after a full iteration step, the current update is rejected, the solution is rolled back to the lowest-residual result from the previous step, and the iteration terminates. If the flow field remains unchanged after the update, the current state is considered a fixed point of the framework, and the iteration also stops. In addition, a maximum iteration count is imposed to prevent infinite recursion.

\section{Numerical Results}

\subsection{Training Performance of the Individual Components}

This subsection first reports the training performance of the individual model components.

From the dataset introduced in Section 2.1, 5\% of the samples are randomly selected as the validation set, and the remaining samples are used for training. Since the inference scenario of the Residual-based Corrector and the Shock-based Corrector is not fully identical to their training setting, the validation set here is used only to assess training performance.

Both correctors are trained for 300 epochs with batch size 64, linear-warm and cosine decay learning rate with max learning rate 2e-04. Their main quantitative differences lie in the masking strategy. Table~\ref{tab:2} summarizes the model sizes, key training settings, and the final validation errors on the in-distribution validation set.

\begin{table}[htbp]
\centering
\caption{The model settings and training performance of the individual components}
\label{tab:2}
\small
\setlength{\tabcolsep}{4pt}
\begin{tabularx}{\linewidth}{@{}ll*{3}{>{\centering\arraybackslash}X}@{}}
\toprule
 & & Flow Initializer & Residual-based Corrector & Shock-based Corrector \\
\midrule
Model settings & Layer number & 8 & 8 & 8 \\
 & Hidden dimension & 256 & 256 & 256 \\
 & Total parameters & 7,021,576 & 7,881,800 & 7,881,800 \\
\addlinespace
Training settings & Epochs & 300 & 300 & 300 \\
 & Batch size & 64 & 64 & 64 \\
 & Learning rate & 2e-04 & 2e-04 & 2e-04 \\
\addlinespace
Training performance & Valid L1 error & 6.38e-04 & 2.13e-04 & 2.22e-04 \\
 & Valid CL MAE & 3.76e-03 & 1.31e-03 & 9.70e-04 \\
 & Valid CD MAE & 7.49e-04 & 1.82e-04 & 1.14e-04 \\
\bottomrule
\end{tabularx}
\end{table}

\subsection{Performance on the OOD Test Set}

This subsection evaluates the improvement brought by the iterative inference framework over the conventional end-to-end prediction on the OOD test set defined in Section 3.2.

\begin{figure}[htbp]
\centering
\includegraphics[width=\linewidth]{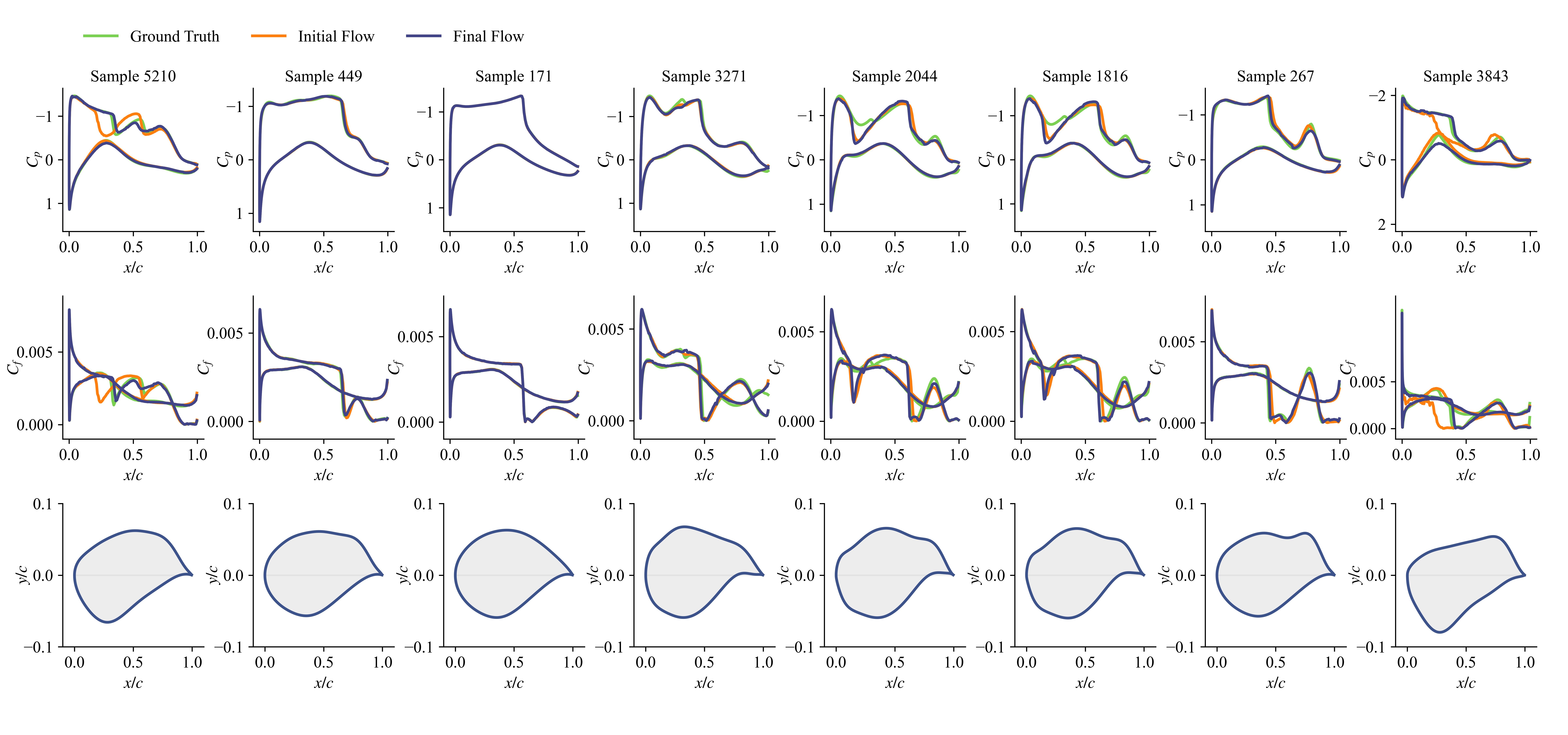}
\par\vspace{4pt}
\includegraphics[width=\linewidth]{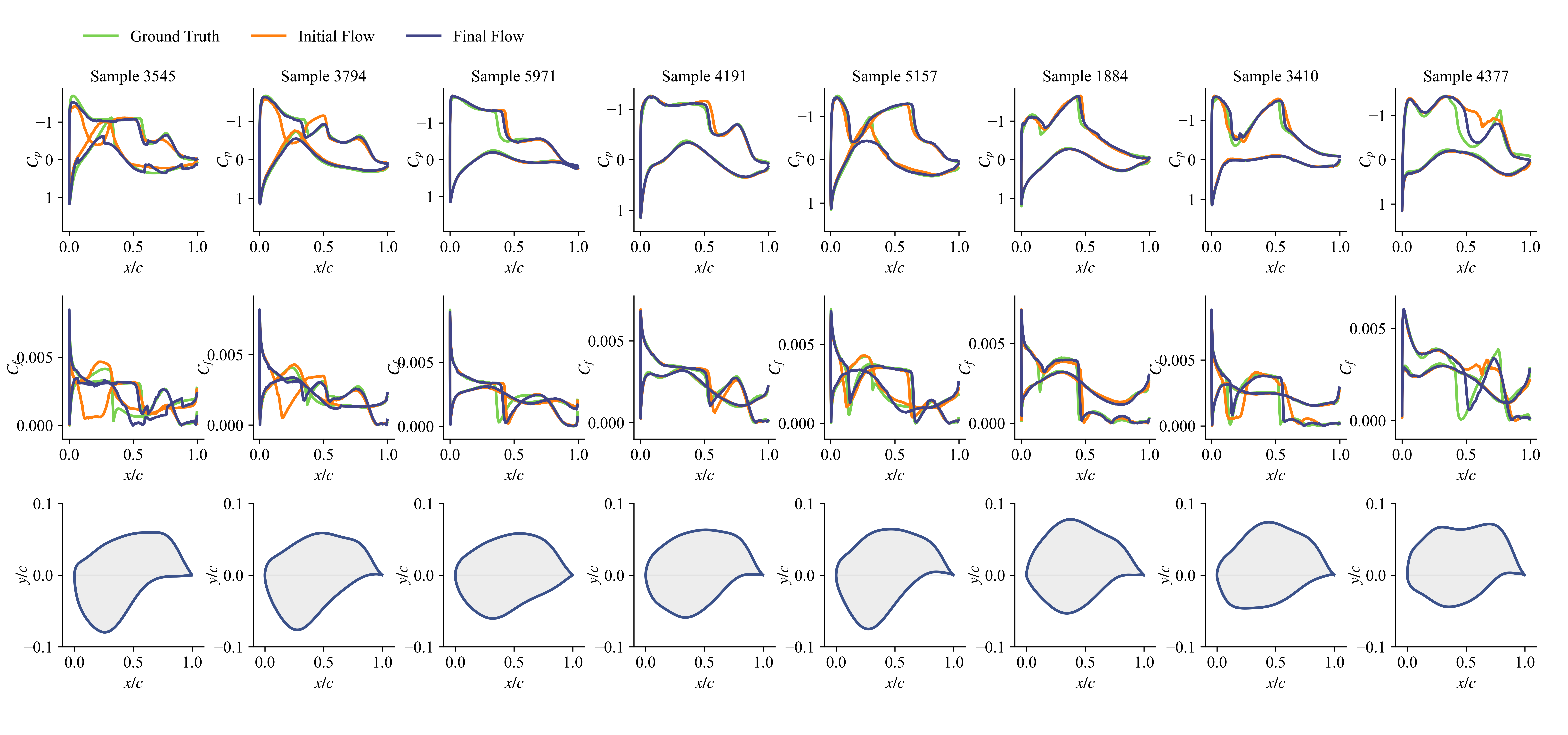}
\caption{The initial and final Cp Cf field of random-selected OOD samples}
\label{fig:4}
\end{figure}

Figure~\ref{fig:4} shows several randomly selected samples from the OOD test set, including the initial flow field predicted by the Geo-to-Flow Flow Initializer, the flow field after iterative inference, and the ground-truth flow field. It can be observed that, because these geometries lie outside the training distribution, the initial flow fields are often unreliable, and important structures such as shocks deviate significantly from the ground truth. After iterative inference, however, the error is substantially reduced. For a few extremely difficult OOD samples, some shock-structure errors still remain even after the iterative framework, while the flow field are significantly more accurate than the initial one.

Table~\ref{tab:3} reports the errors of the initial flow fields on the OOD test set and the errors after applying the iterative framework. The iterative framework reduces the mean L1 error, CL error and CD error on the OOD test set by 30\%, 59\%, and 41\%, respectively. This demonstrates that the proposed framework effectively improves the robustness of the model on out-of-distribution samples.

\begin{table}[htbp]
\centering
\caption{The error of initial and final flow field on the OOD test set}
\label{tab:3}
\small
\begin{tabularx}{\linewidth}{@{}Xrrrr@{}}
\toprule
 & Residual & L1 error & CL MAE & CD MAE \\
\midrule
Initial flow field (single step prediction) & 8.41e-04 & 6.71e-03 & 6.87e-02 & 2.85e-03 \\
Final flow field (multi step iteration) & 4.16e-04 & 4.69e-03 & 2.80e-02 & 1.68e-03 \\
\bottomrule
\end{tabularx}
\end{table}

\begin{figure}[htbp]
\centering
\includegraphics[width=\linewidth]{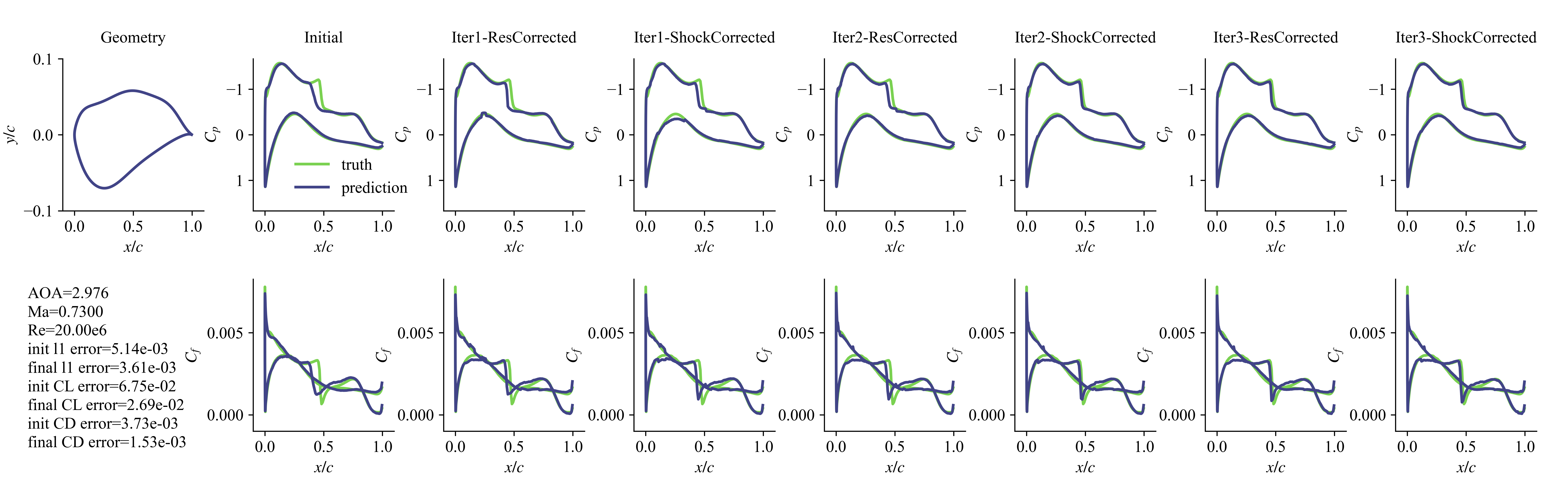}
\par\vspace{4pt}
\includegraphics[width=\linewidth]{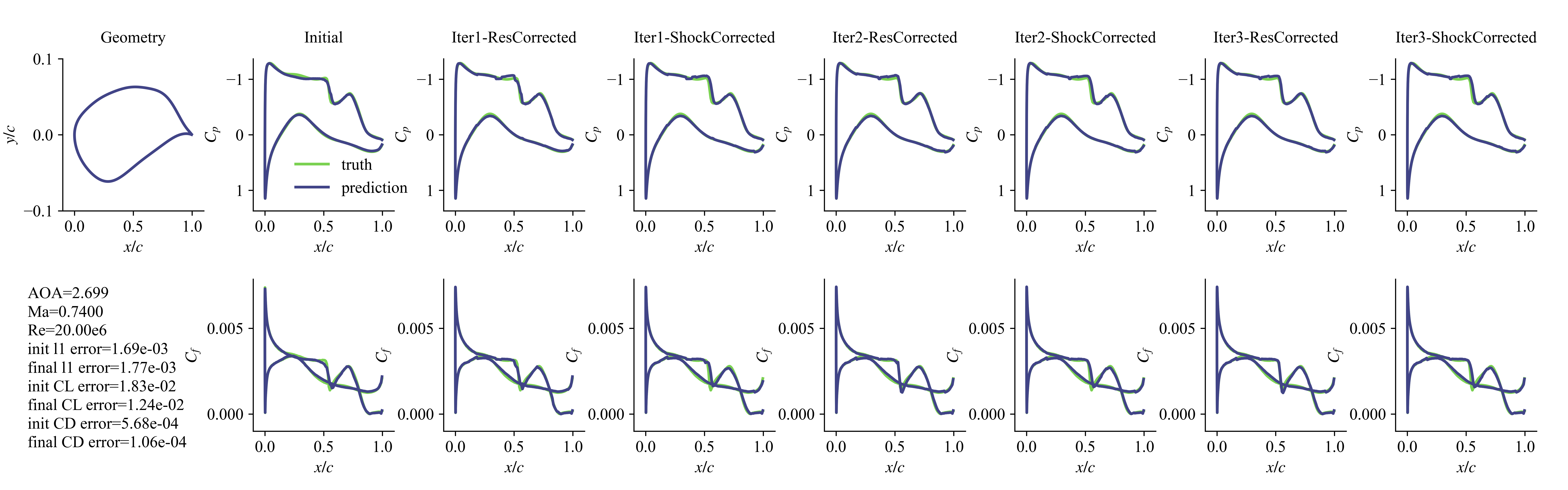}
\par\vspace{4pt}
\includegraphics[width=\linewidth]{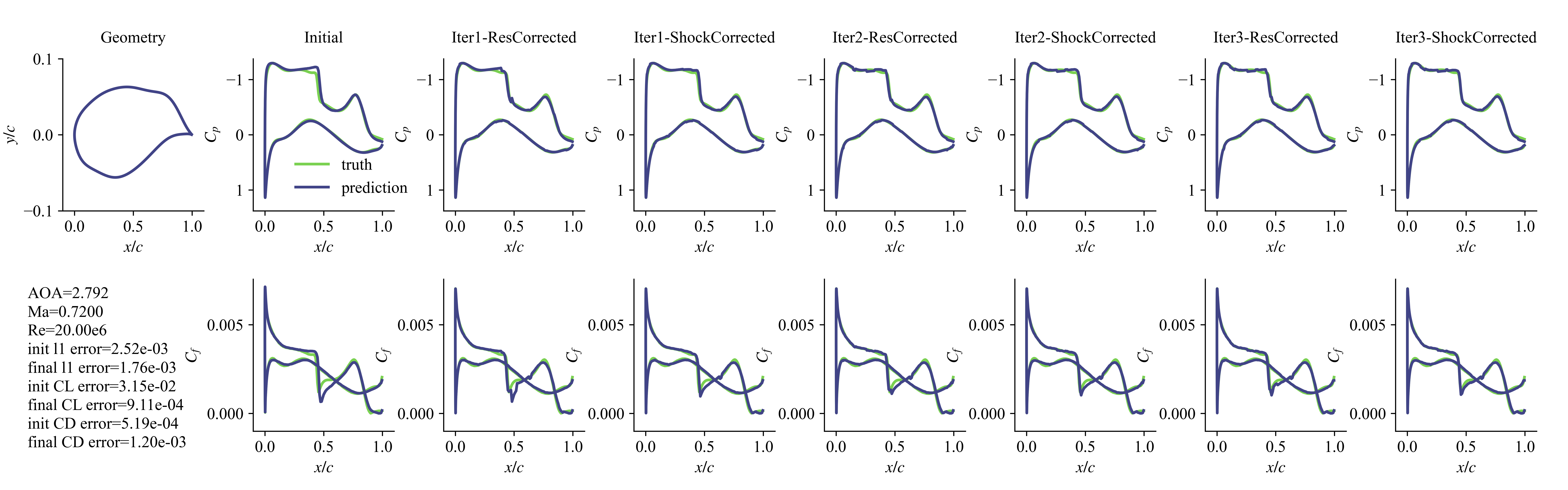}
\caption{The iteration history of random-selected 3 OOD samples}
\label{fig:5}
\end{figure}

Figure~\ref{fig:5} presents the convergence histories of the flow field during the iterative process for three random samples from the OOD test set. All three samples converge after three rounds of alternating-operator iteration, and the flow-field error can be seen to decrease progressively as the iteration advances.

\subsection{Computational Cost}

For conventional Geo-to-Flow prediction, flow-field inference is typically completed within less than 0.1 seconds. Since the proposed framework introduces multi-step operator iteration and residual evaluation, it is necessary to quantify the resulting computational cost.

Table~\ref{tab:4} reports the average iteration count, computational cost, comparison with the Geo-to-Flow model, and the computing device for the OOD test set under batch size 1.

\begin{table}[htbp]
\centering
\caption{The inflow distribution of training and OOD test set}
\label{tab:4}
\small
\begin{tabularx}{\linewidth}{@{}p{45mm}*{4}{>{\centering\arraybackslash}X}@{}}
\toprule
 & Average iteration & Median iteration & Computational cost & Computational device \\
\midrule
Initial flow field (single step prediction) & -- & -- & 0.061s & $1\times$RTX4060 \\
Final flow field (multi step iteration) & 3.47 & 3 & 0.309s & $1\times$RTX4060 \\
\bottomrule
\end{tabularx}
\end{table}

It can be observed that the average number of iterations on the OOD test set is only 3.47, and the average per-sample runtime is only 0.25 seconds. Although this is higher than that of direct Geo-to-Flow prediction, it remains negligible compared with CFD mesh generation and simulation, which highlights the engineering potential of the proposed framework. Further reductions in computational cost are expected once the iterative framework is implemented in a multi-batch parallel setting.

\section{Conclusion and Future Work}

This study constructs a multi-step surface-flow inference framework for improving OOD generalization. The model is trained on limited geometric dataset and tested on an OOD airfoil set collected from real aerodynamic optimization.

Numerical results show that the proposed framework is effective on the optimization-trajectory-based OOD test set. Relative to the usual Geo-to-Flow prediction, the final iterative framework reduces the mean L1 error by 30.07\%, the mean CL error by 59.19\%, and the mean CD error by 41.21\%. At the same time, the additional computational overhead is only about 0.25 s per sample. These results indicate that the framework has clear potential as a practical module for engineering aerodynamic surrogate models.

Future work will focus on extending the current framework from two-dimensional airfoil flows to three-dimensional aerodynamic problems.

\end{document}